\documentclass[aps,pra,showkeys,twocolumn,superscriptaddress]{revtex4-2}
\usepackage[utf8]{inputenc}
\usepackage{amsmath,graphicx,array,float,amsthm}
\usepackage[most]{tcolorbox}
\usepackage{lipsum}
\usepackage[colorlinks={true}]{hyperref}
\hypersetup{colorlinks=true,linkcolor=red,citecolor=blue,urlcolor=blue}
\usepackage[FIGTOPCAP,raggedright,nooneline,bf,footnotesize]{subfigure}
\usepackage{indentfirst}
\usepackage{braket}
\usepackage{diagbox}
\usepackage{amssymb}
\usepackage{bbold}
\usepackage{colortbl}
\usepackage{multirow}
\usepackage{orcidlink}
\usepackage{pstricks}
\begin{document}

 \title{Trade-off relations of quantum resource theory in Heisenberg models} 
\author{Asad Ali \orcidlink{0000-0001-9243-417X}} \email{asal68826@hbku.edu.qa}
\affiliation{Qatar Centre for Quantum Computing (QC2), College of Science and Engineering, Hamad Bin Khalifa University, Qatar Foundation, Doha, Qatar}
\author{Saif Al-Kuwari \orcidlink{0000-0002-4402-7710}} \email{smalkuwari@hbku.edu.qa}
\affiliation{Qatar Centre for Quantum Computing (QC2), College of Science and Engineering, Hamad Bin Khalifa University, Qatar Foundation, Doha, Qatar}
\author{Saeed Haddadi \orcidlink{0000-0002-1596-0763}} \email{haddadi@semnan.ac.ir}
\affiliation{Faculty of Physics, Semnan University, P.O. Box 35195-363, Semnan, Iran}

\begin{abstract}
Studying the relations between entanglement and coherence is essential in many quantum information applications. For this, we consider the concurrence, intrinsic concurrence and first-order coherence, and evaluate the proposed trade-off relations between them. In particular, we study the temporal evolution of a general two-qubit XYZ Heisenberg model with asymmetric spin-orbit interaction under decoherence and analyze the trade-off relations of quantum resource theory.
For XYZ Heisenberg model, we confirm that the trade-off relation between intrinsic concurrence and first-order coherence holds. Furthermore, we show that the lower bound of intrinsic concurrence is universally valid, but the upper bound is generally not.
These relations in Heisenberg models can provide a way to explore how quantum resources are distributed in spins, which may inspire future applications in quantum information processing. 
\end{abstract}
\maketitle

\section{Introduction}\label{sec1}
In quantum resource theory, entanglement and coherence are essential resources to exploit the properties of quantum systems. Entanglement measures non-classical correlations between subsystems, while coherence represents the superposition of systems as a whole. For multipartite systems, quantum coherence is operationally equivalent to entanglement \cite{streltsov2015measuring}. Therefore, it is natural to seek a functional relationship between these two concepts.

For pure two-qubit systems, a complementary relationship exists between first-order coherence (FOC) and concurrence (C) \cite{fan2019universal, dong2023complementary, dong2022unification}. However, this relationship does not hold for mixed states. To address this, researchers in the same work \cite{fan2019universal} have proposed a new quantity, intrinsic concurrence (IC), which exhibits a trade-off with FOC in two-qubit mixed states. Although IC is not generally an entanglement monotone, it becomes equivalent to concurrence for two-qubit pure states. Additionally, the authors in \cite{zhou2020mutual}  have tried to relate IC and concurrence through an inequality relation without providing a rigorous mathematical proof. They claimed that this is true for any arbitrary two-qubit quantum system.  

Spin chains have gained increasing attention within various theoretical and experimental quantum computing platforms, mainly due to their integrability and scalability \cite{meier2003quantum, yu2023simulating, boixo2014evidence, harris2018phase, labuhn2016tunable, kikuchi2023realization}.  
In fact, spin chains have become valuable tools for realizing quantum coherence and correlations, particularly in the context of Heisenberg models \cite{youssef2023exploring, zhang2023comparing, abd2023comparative,verstraelen2023quantum, elghaayda2023quantum, khedif2023intrinsic, ait2023relationship, kuznetsova2023quantum, kuzmak2023entanglement, zhang2023relationship, zhang2018quantum, yurischev2020quantum,park2019thermal,sun2020dynamics, khedif2022non, shi2019entropic,HaddadiEPJP,li2008entanglement, aldosari2023control,YURISCHEV2023128868,Benabdallah2023,universe9010005,MOHAMED2022105693,Hashem105693}. Owing to the various configurations of spin correlation, spin chain models have significant amounts of quantum resources in entanglement and coherence. This has led to their use in quantum computers and quantum networks \cite{yu2023simulating, tacchino2020quantum, wang2022error,Shahbeigi_2022,wu2023qubits, vandersypen2019quantum, ekstrom2022simulating}.
The two-qubit Heisenberg XYZ model is an example of the smallest yet most general spin chain, which has recently motivated the research community to rapidly adopt the Heisenberg XYZ model \cite{youssef2023exploring, zhang2023comparing, abd2023comparative,verstraelen2023quantum, elghaayda2023quantum, khedif2023intrinsic, ait2023relationship, kuznetsova2023quantum, kuzmak2023entanglement, zhang2023relationship, zhang2018quantum, yurischev2020quantum,park2019thermal,sun2020dynamics, khedif2022non, shi2019entropic,HaddadiEPJP,li2008entanglement, aldosari2023control,YURISCHEV2023128868,Benabdallah2023,universe9010005,MOHAMED2022105693,Hashem105693}. 

We, too, are motivated to consider the dissipative two-qubit Heisenberg XYZ under Dzyaloshinsky-Moriya (DM) interaction \cite{mohamed2020quantum, mohamed2021two, abdel2023thermal} and to study the trade-off relations for this model. Furthermore, we test whether the proposed inequality relation holds for this system and analyze the trade-off relation to gain insights into the collective behavior of entanglement and coherence over time for varying system parameters such as purity of initially established state, strength of DM interaction, and damping parameter.

The contribution of this paper is twofold. First, we show that the trade-off relation between IC and FOC holds tightly for the Heisenberg XYZ model. Second, although the lower bound of IC is universally valid, we show that the upper bound is generally not.

The rest of this article is organized as follows: in Sect.~\ref{sec2}, we briefly introduce C, FOC, and IC. We also discuss the trade-off relations between FOC and IC, and the restricted inequality relation between IC and C.  We use the Heisenberg XYZ model to analyze the trade-off relations between C and IC in Sec.~\ref{sec3}.  In Sec.~\ref{sec4}, we present our results and findings, and finally conclude the paper in Sec.~\ref{sec5}.
\section{Preliminaries}\label{sec2}
In this section, we first provide the definitions of IC, C, and FOC. Then, we discuss the trade-off relation between IC and FOC. Finally, we present the inequality relation between IC and C.

For a bipartite system, C gives a measure of entanglement. For two-qubit pure states $|\psi\rangle$, it is defined as 
\begin{equation}
   C(|\psi\rangle) =\sqrt{2\{1-\textmd{tr}[(\rho^A)^2]\}}=\sqrt{2\{1-\textmd{tr}[(\rho^B)^2]\}},\label{eq1}
\end{equation}
where $\rho^{A}=\textmd{tr}_{B}(\rho^{AB})$ and $\rho^{B}=\textmd{tr}_{A}(\rho^{AB})$ are the reduced density matrices of the sub-systems.
For general two-qubit mixed states, C is defined as  \cite{fan2019universal, wooter1998}
\begin{equation}
C(\rho^{AB})=\max\{0,\sqrt{\lambda_{1}}-\sqrt{\lambda_{2}}-\sqrt{\lambda_{3}}-\sqrt{\lambda_{4}}\},\label{eq2}
\end{equation}
where $\lambda_{i}~(i=1,2,3,4)$ are the eigenvalues of non-Hermitian operator $\rho^{AB}\widetilde{\rho}^{AB}$ in the decreasing order. 
 Here, the spin-flipped  density matrix has been defined as $\widetilde{\rho}^{AB}=(\sigma_{y}\otimes\sigma_{y})(\rho^{AB})^{*}(\sigma_{y}\otimes\sigma_{y})$ in which  $(\rho^{AB})^{*}$ is the complex conjugate of $\rho^{AB}$ and $\sigma_{y}$ is the $y$-component of Pauli matrices.
 
If the FOC of each reduced density matrix $\rho^{(A,B)}$ is given by $F_{(A,B)}=\sqrt{2\textmd{tr}[(\rho^{(A,B)})^{2}]-1}$, then the degree of FOC in $\rho^{AB}$ can be evaluated by \cite{fan2019universal, ali2021properties, svozilik2015revealing}
\begin{equation}
F(\rho^{AB})=\sqrt{\frac{F_{A}^{2}+F_{B}^{2}}{2}}.\label{eq3}
\end{equation}

For any arbitrary two-qubit state $\rho^{AB}$, IC can be mathematically expressed as \cite{fan2019universal, dong2023complementary, dong2022unification}
\begin{equation}
C_{I}(\rho^{AB})=\sqrt{\textmd{tr}(\rho^{AB}\widetilde{\rho}^{AB})}=\sqrt{\lambda_{1}+\lambda_{2}+\lambda_{3}+\lambda_{4}},\label{eq4}
\end{equation}
where $\lambda_{4} \leq \lambda_{3} \leq \lambda_{2} \leq \lambda_{1}$. It is pertinent to note that for any pure two-qubit state $\vert \psi\rangle$, IC reduces to C. In other words,  C and IC turn out to be equal $C_{I}(\left\vert \psi\right\rangle )=C(\left\vert \psi\right\rangle )$. In this case, the conservation relation between C and FOC can be written as \cite{fan2019universal, dong2023complementary, dong2022unification}
\begin{equation}
 C^2(\vert \psi\rangle) + F^2(\vert \psi\rangle)=1.\label{eq5}
\end{equation}

Note that this relation does not hold for any two-qubit mixed quantum state. However, a  conservation equation between IC and FOC can be formulated as \cite{fan2019universal}
 \begin{equation}
C_{I}^{2}(\rho^{AB})+F^{2}(\rho^{AB})=\textmd{tr}[(\rho^{AB})^{2}].\label{eq6}
\end{equation}

Interestingly, the conserved quantity is purity $P(\rho^{AB})=\textmd{tr}[(\rho^{AB})^{2}]$.  Moreover, it has also been reported that for any general two-qubit state $\rho^{AB}$, C is the lower bound of IC, which is given by $C(\rho^{AB})\le C_{I}(\rho^{AB})$. The bound is saturated for the case of pure states. 
The authors in \cite{zhou2020mutual} tried to find the maximum upper bound of IC, which is found to be $\sqrt{\frac{1+C^{2}(\rho^{AB})}{2}}$. They proposed that the mutually restricted inequality relation between C and IC for a two-qubit state $\rho^{AB}$ of any rank is
\begin{equation}
C(\rho^{AB})\leq C_{I}(\rho^{AB})\leq\sqrt{\frac{1+C^{2}(\rho^{AB})}{2}}.\label{eq7}
\end{equation}

\section{Theoretical model and system dynamics}\label{sec3}
The Heisenberg XYZ model with added DM interaction can be mathematically described by the following Hamiltonian 
{\small\begin{equation}
H_{S}=\sum_{k=1}^{n}J_{x}\sigma_{k}^{x}\sigma_{k+1}^{x}+J_{y}\sigma_{k}^{y}\sigma_{k+1}^{y}+J_{z}\sigma_{k}^{z}\sigma_{k+1}^{z}+\overrightarrow{D}.(\overrightarrow{\sigma}_{k}\times\overrightarrow{\sigma}_{k+1}),\label{eq11}
\end{equation}}
where  $J_{i}(i=x,y,z)$ represents the strength of spin-spin coupling along the $x$, $y$ and $z$- directions, respectively, $\{\sigma_{k}^{x},\sigma_{k}^{y},\sigma_{k}^{z}\}$ represents the set of Pauli matrices corresponding to the $k$th qubit, and $\overrightarrow{D}$ is DM vector. Here, we take the two-qubit case with a DM vector acting only along the $z$-direction, $D_z\stackrel{\text { def }}{=}\chi$, for which Eq. (\ref{eq11}) becomes
\begin{equation}
H=J_{x}\sigma_{1}^{x}\sigma_{2}^{x}+J_{y}\sigma_{1}^{y}\sigma_{2}^{y}+J_{z}\sigma_{1}^{z}\sigma_{2}^{z}+\chi(\sigma_{1}^{x}\sigma_{2}^{y}-\sigma_{1}^{y}\sigma_{2}^{x}).\label{eq12}
\end{equation}

Diagonalizing the Hamiltonian in Eq. (\ref{eq12}) on two-qubit spin basis $\{\left|\phi_{1}\right\rangle =\left|00\right\rangle ,\left|\phi_{2}\right\rangle =\left|01\right\rangle ,\left|\phi_{3}\right\rangle =\left|10\right\rangle ,\left|\phi_{4}\right\rangle =\left|11\right\rangle \}$ gives the following eigenstates $\left|\psi_{c}\right\rangle$: \begin{equation}
\begin{pmatrix}\left|\psi_{1}\right\rangle \\
\left|\psi_{2}\right\rangle \\
\left|\psi_{3}\right\rangle \\
\left|\psi_{4}\right\rangle 
\end{pmatrix}=\begin{pmatrix}\frac{1}{\sqrt{2}} & 0 & 0 & \frac{1}{\sqrt{2}}\\
0 & \frac{1}{\sqrt{2}} & \frac{\xi}{\sqrt{2}} & 0\\
0 & \frac{1}{\sqrt{2}} & -\frac{\xi}{\sqrt{2}} & 0\\
\frac{1}{\sqrt{2}} & 0 & 0 & \frac{1}{\sqrt{2}}
\end{pmatrix}\begin{pmatrix}\left|\phi_{1}\right\rangle \\
\left|\phi_{2}\right\rangle \\
\left|\phi_{3}\right\rangle \\
\left|\phi_{4}\right\rangle 
\end{pmatrix},
\end{equation}\label{eq13}
and the corresponding eigenvalues $E_{c}$:
\begin{align}
&E_{1}=J_{x}-J_{y}+J_{z}, \quad E_{2}=J_{z}+\beta,\nonumber \\
&E_{3}=J_{z}-\beta,\quad E_{4}=-J_{x}+J_{y}+J_{z}, \label{eq14}
\end{align}
where $\beta=\sqrt{4\chi^{2}+(J_{x}+J_{y})^{2}}$  and $\xi=u+iv$ with $u=(J_{x}+J_{y})/\beta$ and $v=2\chi/\beta$.

Let's assume that our system is a phase-decoherent model where the energy of the two-qubit system is conserved and the reservoir is coupled with two-qubit raising and lowering operators, i.e. $\sigma_{i}^{+}=\left|0_{i}\right\rangle \left\langle 1_{i}\right|$  and $\sigma_{i}^{-}=\left|1_{i}\right\rangle \left\langle 0_{i}\right|$,  where $i=A$ denotes first spin qubit whereas $i=B$ denotes second spin qubit. 
\begin{figure*}[t]
  \centering
  \includegraphics[width=0.75\columnwidth]{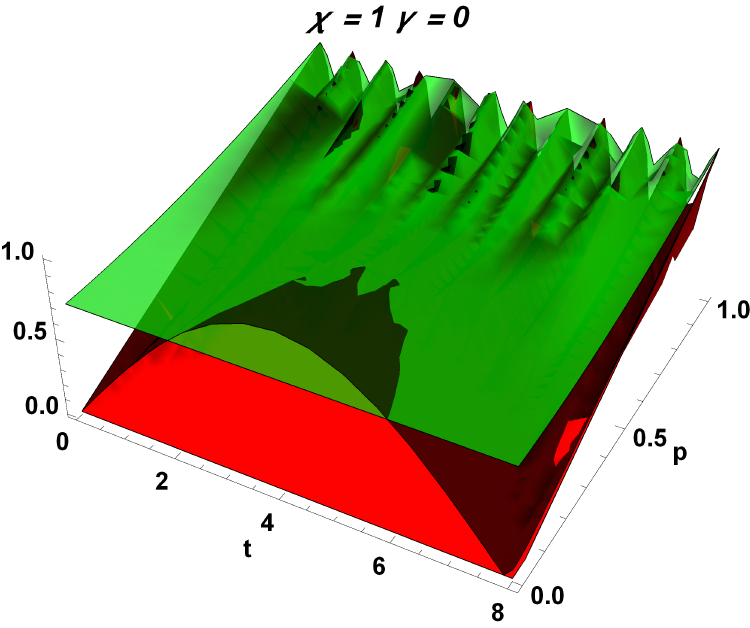} \qquad \quad
   \includegraphics[width=0.75\columnwidth]{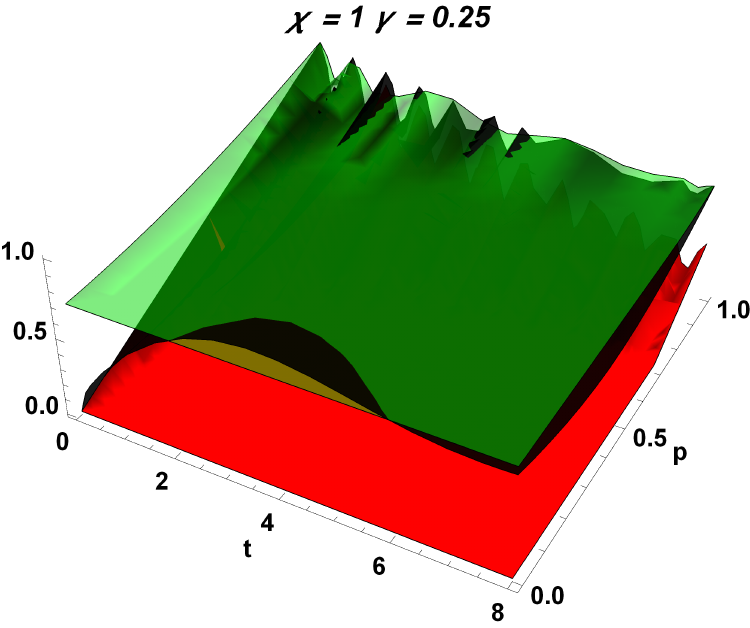}
  \caption{C (red), IC (black), and upper bound of IC (green) in (left) decoherence-free and (right) phase decoherent scenario.}
  \label{f2}
\end{figure*}
The master equation for a given model can be expressed as 
\begin{align}
\frac{d}{dt}\rho^{AB}(t) = & -i[H, \rho^{AB}(t)] \nonumber \\
& + \frac{1}{2}\sum_{i=A,B}\gamma_{i}\big[2\left|1_{i}\right\rangle \left\langle 1_{i}\right|\rho^{AB}(t)\left|1_{i}\right\rangle \left\langle 1_{i}\right| \nonumber \\
& -\left|1_{i}\right\rangle \left\langle 1_{i}\right|\rho^{AB}(t)-\rho^{AB}(t)\left|1_{i}\right\rangle \left\langle 1_{i}\right|\big], \label{eq15}
\end{align}
where $\gamma_{i}$ is the phase damping rate for the $i${th} qubit. For brevity, we assume $\gamma_{A}=\gamma_{B}=\gamma$.

To find the analytical solution for the master equation expressed in Eq. (\ref{eq15}), we consider the situation in which the two-qubit state of the system is initialized in the Horodecki state
\begin{equation}
\rho^{AB}(0)=p\left|\Phi\right\rangle \left\langle \Phi\right|+(1-p)\left|1_{A}1_{B}\right\rangle \left\langle 1_{A}1_{B}\right|,\label{eq18}
\end{equation}
where $0\leq p\leq1$ is the purity parameter of the initial state and $\left|\Phi\right\rangle =\frac{1}{\sqrt{2}}(\left|10\right\rangle +\left|01\right\rangle )$ is Bell state. Then, with the chosen initial state in Eq. \eqref{eq18} and solving the master equation in Eq. \eqref{eq15}, we obtain the non-zero elements of the time-evolved density matrix as
\begin{align}\label{timematrix}
\rho_{11}(t)&=\frac{1}{2}(1-p)[1-e^{-\frac{\gamma t}{2}}\cos2(J_{x}-J_{y}) t],\nonumber\\
\rho_{22}(t)&=\frac{p}{2}-\frac{pv}{2}e^{-\frac{\gamma t}{2}}\sin2\beta t,\nonumber\\
\rho_{33}(t)&=\frac{p}{2}+\frac{pv}{2}e^{-\frac{\gamma t}{2}}\sin2\beta t,\\
\rho_{44}(t)&=\frac{1}{2}(1-p)[1+e^{-\frac{\gamma t}{2}}\cos2(J_{x}-J_{y}) t],\nonumber\\
\rho_{14}(t)&=(\rho_{41}(t))^{*}=-\frac{i}{2}(1-p)e^{-\frac{\gamma t}{2}}\sin2(J_{x}-J_{y}) t,\nonumber\\
\rho_{23}(t)&=(\rho_{32}(t))^{*}={\frac{p \xi}{2}}e^{-\frac{\gamma t}{2}}(u+iv\cos2\beta t).\nonumber
\end{align}

\section{Results}\label{sec4}
In this section, we explore the relations between $C_{I}(\rho^{AB})$, $C(\rho^{AB})$, and $F(\rho^{AB})$ and examine their joint evolution over time. This analysis also involves evaluating $C_{I}(\rho^{AB})^{2}+F^{2}(\rho^{AB})$ across various fixed parameter values, including the purity parameter of the initially established Horodecki-state denoted by $p$, $z-$component of the DM interaction parameter $\chi$, and the phase-damping parameter $\gamma$. In the following, we consider a random case where $J_x=0.5$, $J_y=0.3$, and $J_z=0.8$.

\subsection{IC Bounds Validation}
To validate whether the lower and upper bounds of IC as shown in Eq. (\ref{eq7}) hold, we plot $C(\rho^{AB})$, $C_{I}(\rho^{AB})$, and $\sqrt{\frac{1+C^{2}(\rho^{AB})}{2}}$ in both decoherence-free and phase-decoherent scenarios as shown in Fig. \ref{f2}. 

Fig.~\ref{f2}(a) shows the time evolution of $C(\rho^{AB})$, $C_{I}(\rho^{AB})$, and $\sqrt{\frac{1+C^{2}(\rho^{AB})}{2}}$ as functions of $t$ and $p$ in the decoherence-free scenario at $\chi=1$. It is clear from the figure that $C(\rho^{AB})$ is always below $C_{I}(\rho^{AB})$, confirming that the lower bound of Eq.~(\ref{eq7}) is consistent. 

Then, looking at the green plot representing the upper bound of IC, that is, $\sqrt{\frac{1+C^{2}(\rho^{AB})}{2}}$, we observe that for most parts, $\sqrt{\frac{1+C^{2}(\rho^{AB})}{2}}$ is above $C_{I}(\rho^{AB})$. However, the plot exhibits a few regions where the black plot exceeds the green plot, implying that $\sqrt{\frac{1+C^{2}(\rho^{AB})}{2}} \leq C_{I}(\rho^{AB})$ instead of $C_{I}(\rho^{AB}) \leq \sqrt{\frac{1+C^{2}(\rho^{AB})}{2}}$. Therefore, we conclude that although the lower bound of IC expressed in Eq.~(\ref{eq7}) is universally valid, the upper bound is generally not.

Similarly, to check this bound in the phase-decoherent scenario, we plotted Fig.~\ref{f2}(b) for these functions versus $t$ and $p$ at $\gamma=0.25$ and $\chi=1$, where we can again see that $\sqrt{\frac{1+C^{2}(\rho^{AB})}{2}} \leq C_{I}(\rho^{AB})$ for some specific values of $p$. Therefore, we confirm that the proposed upper bound of IC is not always satisfied. Although the proposed upper bound of IC is valid for $rank-1$ and $rank-2$ density matrices, as proved in \cite{zhou2020mutual}, it is not satisfied for our considered model \eqref{timematrix}, which is a $rank-4$ density matrix.

\begin{figure*}[t]%
\centering
\includegraphics[width=1\textwidth]{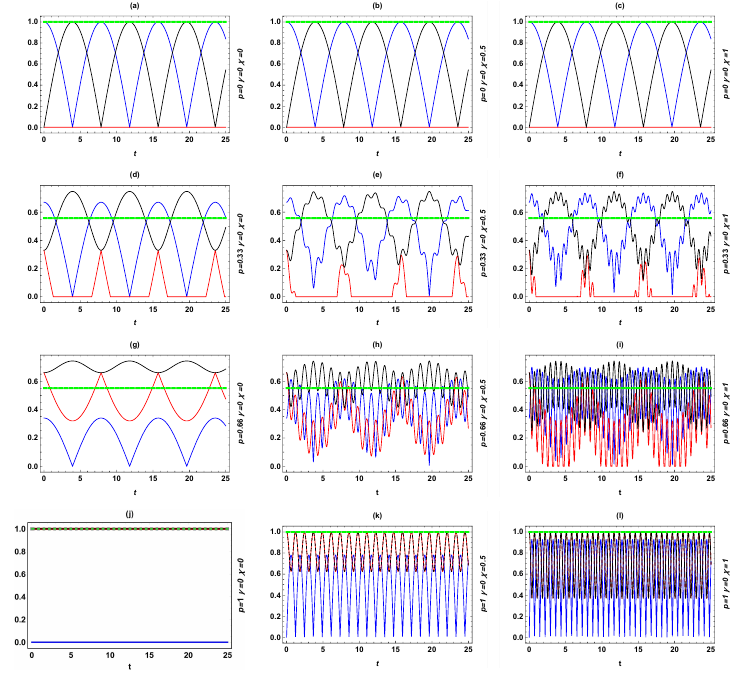}
\caption{The decoherence-free variation of $C(\rho^{AB})$ (red), $F(\rho^{AB})$ (blue), $C_{I}(\rho^{AB})$ (black), $C_{I}^{2}(\rho^{AB})+F^{2}(\rho^{AB})$ (gray), and $P(\rho^{AB})$ (green) as a function of $t$ for different values of $p$ and $\chi$. Note that the curves of $C_{I}^{2}(\rho^{AB})+F^{2}(\rho^{AB})$ (thick-solid gray) and $P(\rho^{AB})$ (thick-dashed green) are perfectly co-coinciding on each other.}\label{f3}
\end{figure*}


\begin{figure*}[t]%
\centering
\includegraphics[width=1\textwidth]{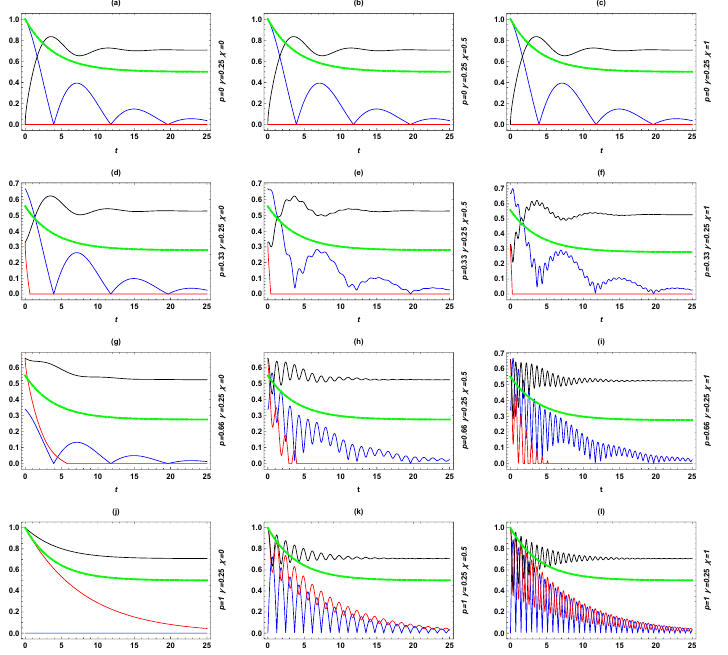}
\caption{The phase-decoherent dynamics of $C(\rho^{AB})$ (red), $F(\rho^{AB})$ (blue), $C_{I}(\rho^{AB})$ (black), $C_{I}^{2}(\rho^{AB})+F^{2}(\rho^{AB})$ (gray), and $P(\rho^{AB})$ (green) as a function of $t$ for different values of $p$ and $\chi$ with $\gamma=0.25$}
\label{f4}
\end{figure*}

In the context of the decoherence-free scenario $\gamma=0$, we now turn our attention to examining the impact of the DM interaction on the previously mentioned quantifiers and the associated complementary trade-off relationship. 
Fig. \ref{f3} shows the collective decoherence-free dynamics of $C(\rho^{AB})$, 
 $F(\rho^{AB})$, $C_{I}(\rho^{AB})$, and $C_{I}^{2}(\rho^{AB})+F^{2}(\rho^{AB})$  as a function of $t$ for different fixed values of $\chi$ and $p$.
 
In Fig. \ref{f3}(\textbf{a}-\textbf{c}), where $p=0$, it can be deduced that the initially separable pure state at $t=0$ maintains its separability for all subsequent times $t>0$. Consequently, C remains consistently zero throughout time. Additionally, our observations reveal that augmenting the strength of the DM interaction, as depicted in Fig. \ref{f3}(\textbf{a})($\chi=0$), Fig. \ref{f3}(\textbf{b})($\chi=0.5$), and Fig. \ref{f3}(\textbf{c})($\chi=1$), does not yield entanglement generation. Furthermore, it is established that the behaviors of entanglement (captured by C) and coherence (captured by FOC) remain consistent regardless of the DM interaction strength at any given time, indicating that the DM interaction cannot induce or enhance either entanglement or coherence in our considered system when $p=0$.

In Fig. \ref{f3}(\textbf{d}-\textbf{f}), where $p=0.33$, 
we notice that, in general, C is non-zero, exhibiting entanglement deaths and revivals over time. As the value of the DM interaction increases, we observe that the DM interaction generates an extra wiggling or ripple effect not only in C but also in IC and FOC. 

Similar behavior is also observed for $p=0.66$ in Fig. \ref{f3}(\textbf{g}-\textbf{i}), but now entanglement is mostly non-zero at $p=0.66$ over time, and the DM interaction, although it cannot increase the maximum value of entanglement, it decreases the minimum value of concurrence. Furthermore, for a larger value of the DM interaction, C tries to mimic the variation of IC, though not exactly with the increase in the value of $p$. In addition, increasing the DM interaction also increases the average oscillation frequency of the C, FOC, and IC. 

Finally, in Fig. \ref{f3}(\textbf{j}-\textbf{l}), the initially prepared Horodecki state becomes the Bell state at $p=1$. We observe that C and IC behave identically and that increasing the value of the DM interaction significantly increases the oscillation frequency. Therefore, We can conclude that, in general, increasing the strength of the DM interaction is responsible for increasing the oscillation frequency of FOC, C, and IC, provided that the initially established state is the entangled state. It is also clear that the lower bound of IC, as mentioned in Eq. \eqref{eq7} holds perfectly. The trade-off relation between FOC and IC in the form of a recently proposed conservation relation, i.e. $C_{I}^{2}(\rho^{AB})+F^{2}(\rho^{AB})=Tr[(\rho^{AB})^{2}]$ is valid as the curves of $C_{I}^{2}(\rho^{AB})+F^{2}(\rho^{AB})$ (thick-solid gray) and $P(\rho^{AB})$ (thick-dashed green) coincide perfectly with each other.

\subsection{Phase-decoherent Dynamics}
We now focus on examining the phase-decoherent dynamics of $C_{I}(\rho^{AB})$, $C(\rho^{AB})$, $F(\rho^{AB})$, and $C_{I}^{2}(\rho^{AB})+F^{2}(\rho^{AB})$ over time, while taking varying fixed values of the DM interaction and $\gamma=0.25$, as illustrated in Fig. \ref{f4}. 

As seen in Fig. \ref{f4}(\textbf{a}-\textbf{c}), it is clear that even with non-zero decoherence, an increase in the DM interaction value does not lead to the generation of entanglement or coherence at $p=0$. In contrast to Fig. \ref{f3}(\textbf{d}-\textbf{f}), where entanglement death and revival are observed, Fig. \ref{f4}(\textbf{d}-\textbf{f}) shows that no such phenomenon occurs due to the presence of non-zero decoherence. However, very few entanglement death-revival is evident for relatively larger values of $p$ and larger values of the DM interaction at initial times, as shown in Fig. \ref{f4}(\textbf{g}-\textbf{i}). Similarly, the duration of entanglement survival decreases before the complete entanglement death for larger values of $p$. Furthermore, the trade-off relation is completely held tightly even in the case of non-zero decoherence, including the lower bound of IC as specified in Eq. \eqref{eq7}.
Eventually, in Fig. \ref{f4}(\textbf{j}-\textbf{k}), one can observe that the initialized state is maximally entangled and, therefore, at $t=0$, IC and C have the same value. However, introducing decoherence makes them different and decreases their values over time. Figs. \ref{f3} and \ref{f4} illustrate the importance of studying decoherence effects and initial state parameters to control their destructive effects.

\section{Conclusion}\label{sec5}
The complementarity relation between intrinsic concurrence (IC) and first-order coherence (FOC) in a two-qubit system can shed light on the connection between entanglement and coherence. The trade-off relation, which is universal for any arbitrary two-qubit states, shows that as the level of IC increases, FOC decreases and vice versa. The relationship between IC and FOC is crucial for understanding the transfer of quantum resources, such as entanglement and coherence, between bipartite quantum systems. Moreover, it has been mathematically proven that IC contains the concurrence (C) of four pure states consisting of a special pure state ensemble concerning an arbitrary two-qubit state, and an inequality has been proposed between these measures. Although the upper bound of this inequality has yet to be proven, the complementarity relation between IC and FOC provides a valuable tool for understanding the complex behavior of quantum systems.

In this work, we studied the dynamics of IC, FOC, and C as a function of time with and without phase decoherence, assuming the initial state to be a Horodecki state based on maximally entangled Bell state and interaction Hamiltonian of two-qubit Heisenberg XYZ model under Dzyaloshinsky-Moriya (DM) interaction. We found that the DM interaction plays a vital role in the behavior of the IC, FOC and C in both decoherent and decoherence-free scenarios. We also checked the validity of the trade-off relation between IC and FOC and the inequality relation between IC and C. We confirmed that in the given model, the trade-off relation, that is $C_{I}^{2}(\rho^{AB})+F^{2}(\rho^{AB})=Tr[(\rho^{AB}]^{2}$ between IC and FOC holds tightly. However, the upper bound of the inequality between IC and C, namely $C_{I}(\rho^{AB})\leq\sqrt{\frac{1+C^{2}(\rho^{AB})}{2}}$ generally does not hold.

\vspace{1cm}
\section*{Acknowledgements}
 S.H. was supported by Semnan University under Contract No. 21270.
 
\section*{Disclosures}
The authors declare that they have no known competing financial interests.

\section*{Data availability}
No datasets were generated or analyzed during the current study.

\bibliography{bibliography} 
\end{document}